\documentclass[conference]{IEEEtran}
\IEEEoverridecommandlockouts
\usepackage{cite}
\usepackage{amsmath,amssymb,amsfonts}
\usepackage{algorithmic}
\usepackage{graphicx}
\usepackage{textcomp}
\usepackage{xcolor}
\def\BibTeX{{\rm B\kern-.05em{\sc i\kern-.025em b}\kern-.08em
    T\kern-.1667em\lower.7ex\hbox{E}\kern-.125emX}}
\begin{document}

%
\title{From Group Psychology to Software Engineering Research to Automotive R\&D: Measuring Team Development at Volvo Cars}

\author{
\IEEEauthorblockN{Lucas Gren}
\IEEEauthorblockA{\emph{Department of Computer Science and Engineering}\\\emph{University of Gothenburg}\\Gothenburg, Sweden\\
Email: lucas.gren@cse.gu.se}
\and
\IEEEauthorblockN{Christian Jacobsson}
\IEEEauthorblockA{\emph{Department of Psychology}\\\emph{University of Gothenburg}\\Gothenburg, Sweden\\
Email: christian.jacobsson@psy.gu.se}

}

\maketitle

\begin{abstract}
From 2019 to 2022, Volvo Cars successfully translated our research discoveries regarding group dynamics within agile teams into widespread industrial practice.  We wish to illuminate the insights gained through the process of garnering support, providing training, executing implementation, and sustaining a tool embraced by approximately 700 teams and 9,000 employees.  This tool was designed to empower agile teams and propel their internal development.  Our experiences underscore the necessity of comprehensive team training, the cultivation of a cadre of trainers across the organization,  and the creation of a novel software solution.  In essence, we deduce that an automated concise survey tool, coupled with a repository of actionable strategies, holds remarkable potential in fostering the maturation of agile teams, but we also share many of the challenges we encountered during the implementation. 
\end{abstract}

\IEEEpeerreviewmaketitle
\section{Motivation}\label{sec:life}
From 2013 until 2018, we conducted novel research on the connections between the group psychological aspect of team development and the emerging and targeted dynamics of agile teams. In 2018, the first author switched fields from academia to becoming a change leader in the agile transformation at Volvo Cars.  
This article explains the experience gained and the details of how to transfer this knowledge to a large automotive company.

More specifically, Volvo Cars implemented a short version of Wheelan's Group Development Questionnaire \cite{gren2020gdqs} with automated feedback to teams. This paper comprises an overview of our experience of the training, implementation, and maintenance of that tool. 

\section{Background}\label{sec:background}

Wheelan's Integrated Model of Group Development (IMGD) builds upon earlier models by Tuckman \cite{tuckman} and outlines four stages that small groups typically progress through: (1) Dependency and Inclusion (as described by Wheelan \cite{wheelan}) --- Members rely on the leader and focus on inclusion. The goal is to establish safety and structure for work to begin. Disagreements are rare, (2) Counter-dependency and Fight --- With increased safety, members express differing viewpoints. The group works to integrate these and lessen leader dependence. Difficulty integrating differences can lead to scapegoating. The goal is to resolve task conflicts, avoid personal attacks, and clarify goals and procedures. (3) Trust and Structure --- Successfully integrating differences and redistributing authority lead to Stage 3. Trust increases as diverse viewpoints are accepted. Members recognize interdependence. The focus shifts to balancing autonomy with interdependence, and further clarifying goals, roles, and processes, and (4) Work and Productivity --- Continued work on roles, goals, and processes allows the group to reach Stage 4, marked by high effectiveness, cohesion, and work satisfaction. The group functions as a team with minimal leader dependence.

Wheelan \cite{wheelan} created a questionnaire (The Group Development Questionnaire) that captures the team maturity level. There are four Scales corresponding to a development stage comprising 15 items per scale. It was this measurement that was used to show the connection between team maturity and agility in \cite{gren2019agile,gren2019perceived}. The problem was that it had 60 items in total, which is time-consuming to answer with the shorter time-intervals that Volvo Cars wanted in-between measurements. However, a validated short version of the tool has been created, called the GDQS \cite{gren2020gdqs}. This tool comprises 13 items that teams can fill out and obtain a valid measurement of their maturity level.  Volvo Cars was allowed to use these 13 items because there was an ongoing research collaboration with GDQ Associates\footnote{www.gdqassoc.com}.

\paragraph{Using a short questionnaire in practice}
A validated such short version of a longer questionnaire has its advantages (like enabling fast assessment), but also its caveats (mainly removing most of the resolution one gets from having many items). The latter implies that teams should not only look at the 13 items included and try to optimize their mean values over time, since that will likely misguide them and has a high risk of teams missing aspects of group dynamics that they need to work on. To address this, we did not let teams follow mean values of specific questions, but instead only the validated factors of the four scales. The tool, instead, output a team maturity profile overall. Teams then needed to assess all the aspect of their fitted stage by revisiting the Integrated Model of Group Development \cite{wheelandev}. 

\section{Implementation}\label{sec:implementation}
We first tried to find an Off-the-Shelf solution for embedding the survey. However, no technical solution could conduct our needed calculations, display the result, nor protect individuals and teams the way we wanted. We, therefore, implemented our own solution. This solution let a team member (the chosen administrator) set up surveys and generate anonymous codes that team members use to fill in the survey. The administrator in the team could set how many that should answer the survey every time and could also choose to close the survey without a collected answer from everyone. We then collected anonymous codes for teams and their surveys in connection to their specific organizational unit. We could also show anonymous data of all the teams within a unit. We did not tag individuals with anonymous codes, but only collected individual answer through the survey code and used all those answer to calculate the team's result.

We leveraged normative data derived from a comprehensive research dataset comprising 2,600 distinct teams, as reported in the work by Gren et al. ~\cite{gren2020gdqs}. It is pertinent to note that these teams were entirely unrelated to the domain of Volvo Cars.

We provided guidance to these teams by introducing a systematic framework for directing their efforts. This guidance included the presentation of a developmental toolbox,  categorized into the four distinct developmental stages. One example for a team in Stage 1 would be to workshop the clarification of the team's goals and and the team members' roles. As teams progressed in their endeavors, we offered them a view of trend data to facilitate continuous assessment of their developmental maturity alongside their current developmental stage.



\begin{figure*}
    \includegraphics[width=\linewidth]{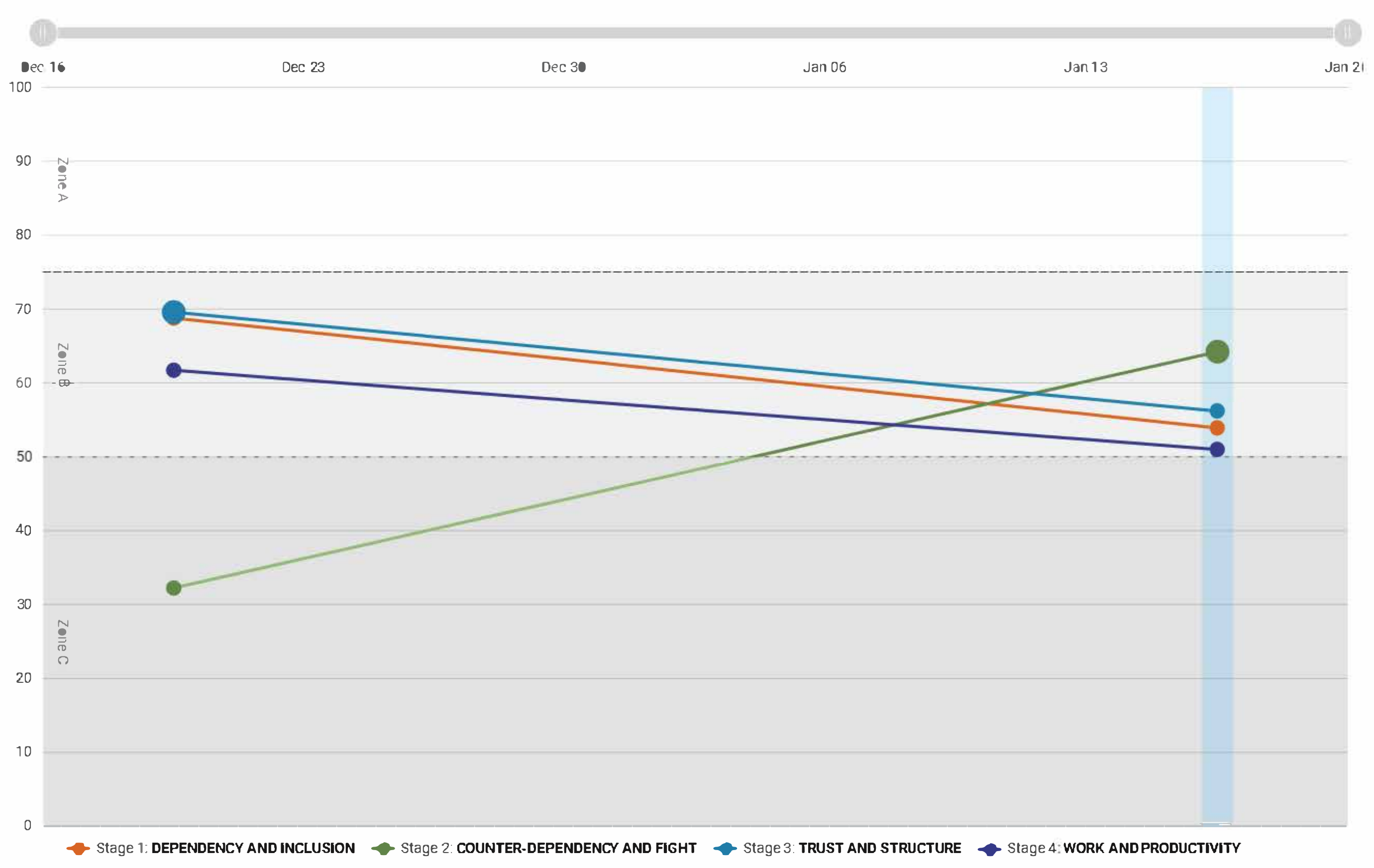} 
    \caption{An example of a team's trend view with two measurements.} 
    \label{example2} 
\end{figure*}

Figure~\ref{example2} is an example of a trend view for a team. On top is the timeline describing when the surveys were completed. The orange dot represents the mean value for Stage 1, the green one Stage 2, the blue one Stage 3 and the purple one Stage 4. The $x$ axis is then time (top bar) and the $y$ axis is a comparison with the norm data, i.e.\ we compare the team's specific result to that data to say if the values are high or low. The 50\% dashed line in the middle is the mean value in the norm data for that stage. This means that the number behind it is different for all the four stages but here transformed into 0 to 100\% in order to not have different plots for each stage. As an example, an orange dot on the 50\% dashed line is a different absolute mean value as compared to a blue dot on the same dashed line. What is relevant is how you score in comparison to the ground truth, not the exact number. 

One challenge when implementing this new short version is that we do not know how far from the norm data mean value a team should be in order to get a match. The top dashed line is one standard deviation from the norm data mean values. A team should have one stage over that line to obtain a clear match in a stage. Under that line, in the light gray area, the team will get a match on the highest mean value in that area, if no value is in Zone A. This was implemented because too many team did not get any stage match, which gave them a hard time to know what to start improving. However, the stage match is just the first step and following the trends and analyzing the difference is what is the most useful. Over time when a team matures from being newly formed, it first get high values on the orange dot, and low on the other ones, then the green dot is high and the team goes through Stage 2. Eventually the orange and green dots go down while the purple and blue ones go up. 

However, it is noteworthy that real-world teams are seldom constituted with entirely new members; pre-existing relationships and dynamics often come into play. In such cases, the interpretation of developmental charts becomes more intricate, especially when relying on a limited number of measurements. Consequently, monitoring and tracking developmental trends over multiple measurements emerge as a valuable strategy, as this approach facilitates a deeper understanding of a team's trajectory and can reveal instances where a need for revisiting previous developmental stages arises. It is imperative to acknowledge that such revisits are often a recurrent necessity, as teams tend to exhibit dynamic and evolving characteristics over time.

In the example (Figure~\ref{example2}), we see a team where something happened the resulted in interpersonal conflict. It could have been the loss of a team member to another team or anything that would get people to sort something out. Training in conflict resolution and how to communicate your standpoint at the same time as making sure the counterpart does not lose face take practice and is most often an effort well spent.  

\paragraph{The Software Tool}

Since this is a quite sensitive assessment, a key that was in focus during the creation and implementation of this tool was a technical environment that was safe for teams. This meant that we handled the data from the measurement according to the principle that, if you helped generate data, you co-owned that data with your team members. When more than three team members filled out the survey, they could see the team's result, not any individual responses. The team also owned their result and showed it to whoever they wanted. The sets of teams (or ARTs) saw all the results for the teams anonymized, meaning that they saw the results from teams $1, 2, 3, ..., n$ but they did not know which team was which.

We also created a web-page on the intranet with guidance for team support (both with tools, workshops, and external help from HR etc.) so that the teams could get the support they needed. The purpose was to create a measurement that helped the teams' internal development. A second useful output was, of course, that the R\&D level could see a all the teams (also anonymized) and where they were in their development. We could there-through eventually understand the ecosystem and investigate why departments did, or did not, obtain self-organized agile teams on a large scale. The focus was to empower the teams. In doing so, we tried to put the team in the driver's seat of their own collaborative maturity, so to speak.

\paragraph{Training}
As a part of empowering teams and increasing their likelihood of becoming autonomous, we trained all teams that want to use the tool before they were given access. Through this, we increased their ability to interpret their results together in the teams and take an active part in improving the team. This was, of course, time-consuming and we were initially a team of around 10 people who met around 600 teams for three hours per team (not all of them had started using the tool, though). In order to increase our training capacity, some of the training was conducted in very large lecture halls with around 250 employees at a time. We also ran a train-the-trainer program where employees from all organizational levels could become trainers and offer the training to teams they were not connected to, thus maintaining the outside perspective of the trainer.

The agile transformation, and this team maturity training as a part of it, started at the Vehicle Software and Electronics department in 2017. The reason was that the software development of the new cars had exploded in size in the last years and the agile and team ideas had reached the farthest at that department. Through the research of Gren \cite{grenphd}, we created and argued for the agile case and its dependence on mature teams. Introducing what the core of agile development is and why empowered and autonomous teams are a key enabler of responsiveness to change \cite{responsiveness}, is an identified success factor in getting so many teams to want to use the tool.

\paragraph{Team development and virtual teams}
Additionally, we provided guidelines for the effective utilization of the tool in virtual teams, drawing inspiration from the work of Hertel et al.~\cite{hertel2005managing}. This was essential since the COVID-19 struck the world during this work.  

In essence, we recommended that the fundamental principles for team formation remained consistent with conventional practices. However, it was imperative to underscore that the establishment of remote teams demands an even greater degree of effort and dedication from the individuals involved. Any challenges encountered in team building are magnified when working in a remote context.

Creating a new team in a remote setting, particularly when team members have never met in person, presents substantial difficulties. Fortunately, many teams at Volvo Cars had prior experience working face-to-face, which facilitates the transition to remote collaboration. We strongly advised that teams enforced a more stringent adherence to their established work practices, given the absence of informal interactions, such as small talk by the coffee machine. Teams should endeavor to make virtual work as closely aligned with regular work as feasibly possible.

Our recommendations aim to empower teams to adapt effectively to remote work, maintain their productivity, and uphold their commitment to collaborative development, all while harnessing the full potential of the tool.

\paragraph{The role of management}
Management should play a supportive role by creating a good environment for teams to grow. Sometimes, management wanted to use team development levels as performance indicators (KPIs). This can be useful, especially if the data is used anonymously to help develop better team-supportive strategies. However, it is important to be careful with this approach. Too much control can make teams just try to meet management’s expectations instead of honestly showing their true progress.

We suggested KPIs that encourage teams to use specific tools and set achievable goals to improve their teamwork. We also recommend avoiding KPIs that unfairly blame teams, such as penalizing them for a member leaving.While open communication and data sharing are ideal in companies, it is hard to achieve, especially in big corporations. Real change needs to come from within, not imposed from the top.Our strategy included educating higher-level managers and encouraging them to use our development tools with their own management teams. This approach worked well, even with senior R\&D managers. 

\section{Lessons learned}\label{sec:lessons}

An important reflection is that we created trust in the way this tool was implemented, since trust is absolutely fundamental in creating empowered teams. We obtained many requests from managers who wanted to see the results from ``their'' teams. Telling all mangers at all levels that they could not see individual teams' results without asking their permission, but only anonymous results from more than three teams was an important part of the agile transformation. The teams used this tool for their internal development only. Then we could look at larger chunks of data to understand where to support teams better, but without any flavors of supervision. The organization seemed to have accepted that we only had one shot at this, and if we misused this data even once, the teams would never answer the survey honestly again. We believe that Volvo Cars did not need a tool to control and monitor teams as much as it needed a tool that empowers them.

In our practical experience, we have observed that teams, when equipped with a thorough understanding of the group development stages, are typically adept at self-assessing accurately when guided through these stages. The tool we have implemented offers a valuable resource by enabling teams to compare their specific stage values with those of numerous other teams in our extensive database. It serves the dual purpose of indicating whether their stage values are relatively high or low across the four developmental stages and aiding in the evaluation of their progress in team development over time.

Furthermore, the tool has made a significant contribution to the development of a company-wide comprehension of the interconnectedness between team maturity and broader performance metrics. This holistic perspective has been instrumental in guiding cultural transformation efforts and other training initiatives within the organization.

Our approach to quantifying psychological constructs, particularly within the domain of social psychology, has been characterized by a nuanced stance. While numerical metrics can provide a level of clarity and accessibility, there is always the inherent risk of oversimplifying complex phenomena to the point where they lose their practical applicability in real-world contexts. Navigating the intricate dynamics of group development when assembling a team is a challenging undertaking, necessitating substantial effort and time.


The strategic shift in helping teams to autonomy, was motivated by our aspiration to cultivate a culture where team members take ownership of their development and engage in continuous learning. Achieving excellence as a team player is an iterative process that often requires dedicated support. Striking a balance between simplifying the measurement to cater to a wide user base, thereby increasing the likelihood of team improvement, and introducing a more complex perspective on collaborative work, which may risk demotivation, was a delicate endeavor. 

Building effective teams, much like other organizational endeavors, is a formidable undertaking. It is labor-intensive work that sometimes takes a deeply personal turn. The introduction of the tool often instilled a sense of hope among employees that they could resolve team-related issues on their own. However, it is important to underscore that the tool cannot act as a panacea for such challenges. Instead, it prompted team members to engage in introspection and contemplate the prerequisites for a well-functioning team. This dialogue permeated various levels of the company, fostering valuable discussions. Notably, many engineers were previously unaware of the existence of a group process before the tool's implementation.

\bibliographystyle{IEEEtran}

\bibliography{ref}

\end{document}